\documentclass[twocolumn]{aastex631}

\usepackage{multirow, gensymb}

\shorttitle{Alignment of Outflows in Serpens Main}
\shortauthors{Green et al.}

\begin{document}

\title{Why are (almost) all the protostellar outflows aligned in Serpens Main?}%
\correspondingauthor{Joel D. Green}
\email{jgreen-at-stsci-dot-edu}

\author[0000-0003-1665-5709]{Joel D. Green}
\affiliation{STScI, 3700 San Martin Dr, Baltimore, MD 21218, USA}

\author[0000-0001-7552-1562]{Klaus M. Pontoppidan}
\affiliation{Jet Propulsion Laboratory, California Institute of Technology, 4800 Oak Grove Drive, Pasadena, CA 91109, USA}

\author[0000-0002-3887-6185]{Megan Reiter}
\affiliation{Rice University, Houston, TX, USA}

\author[0000-0001-8302-0530]{Dan M. Watson}
\affiliation{University of Rochester, Rochester, NY, USA}

\author[0000-0003-0281-7383]{Sachindev S. Shenoy}
\affiliation{STScI, 3700 San Martin Dr, Baltimore, MD 21218, USA}

\author{P. Manoj}
\affiliation{Tata Institute of Fundamental Research, Mumbai, Maharashtra, IN}

\author{Mayank Narang}
\affiliation{Institute of Astronomy and Astrophysics Academia Sinica, Taiepei, TW}


\begin{abstract}

We present deep 1.4-4.8\,$\mu$m JWST-NIRCam imaging of the Serpens Main star-forming region and identify 20 candidate protostellar outflows, most with bipolar structure and identified driving sources. The outflow position angles (PAs) are strongly correlated, and aligned within $\pm 24\degree$ of the major axis of the Serpens filament. These orientations are further aligned with the angular momentum vectors of the two disk shadows in this region. We estimate that the probability of this number of young stars being co-aligned if sampled from a uniform PA distribution is $10^{-4}$. This in turn suggests that the aligned protostars, which seem to be at similar evolutionary stages based on their outflow dynamics, formed at similar times with a similar spin inherited from a local cloud filament. Further, there is tentative evidence for a systematic change in average position angle between the north-western and south-eastern cluster, as well as increased scatter in the PAs of the south-eastern protostars. SOFIA-HAWC+ archival dust polarization observations of Serpens Main at 154 and 214\,$\mu$m are  perpendicular to the dominant jet orientation in NW region in particular. We measure and locate shock knots and edges for all of the outflows and provide an identifying catalog. We suggest that Serpens main is a cluster that formed from an isolated filament, and due to its youth retains its primordial outflow alignment.

\end{abstract}


\section{Introduction} \label{sec:intro}

Star formation is thought to be partly regulated by magnetic fields with coherence scales of a few parsec \citep{Crutcher12} -- smaller than Giant Molecular Clouds, but larger than individual protostars. Magnetic fields likely play a key role in the collapse of cloud cores distributed in elongated structures called filaments \citep{Bally87,Smith16}. Star-forming cores are indeed found to cluster along filamentary density enhancements \citep{Andre10}, however, observational confirmation of a direct influence of the magnetic field has been elusive and there is no consensus on the detailed formation mechanism of filaments and their related young clusters \citep{Hennebelle12, Gomez18}. While theory often assumes idealized alignment of protostellar disks, cores, and associated magnetic field \citep{Konigl00}, feedback may lead to misalignment on the smallest scales (1000\,au) as the protostar evolves \citep{Hull13}. One potential tracer of the accretion flow history of star-forming filaments and their cores on parsec scales is whether the angular momentum vectors of stars in a cluster are correlated with each other, and with direction of the magnetic field along their natal cloud filament \citep{Nagai98}. 

\begin{figure*}[ht!]
    \centering
    \includegraphics[width=0.8\textwidth]{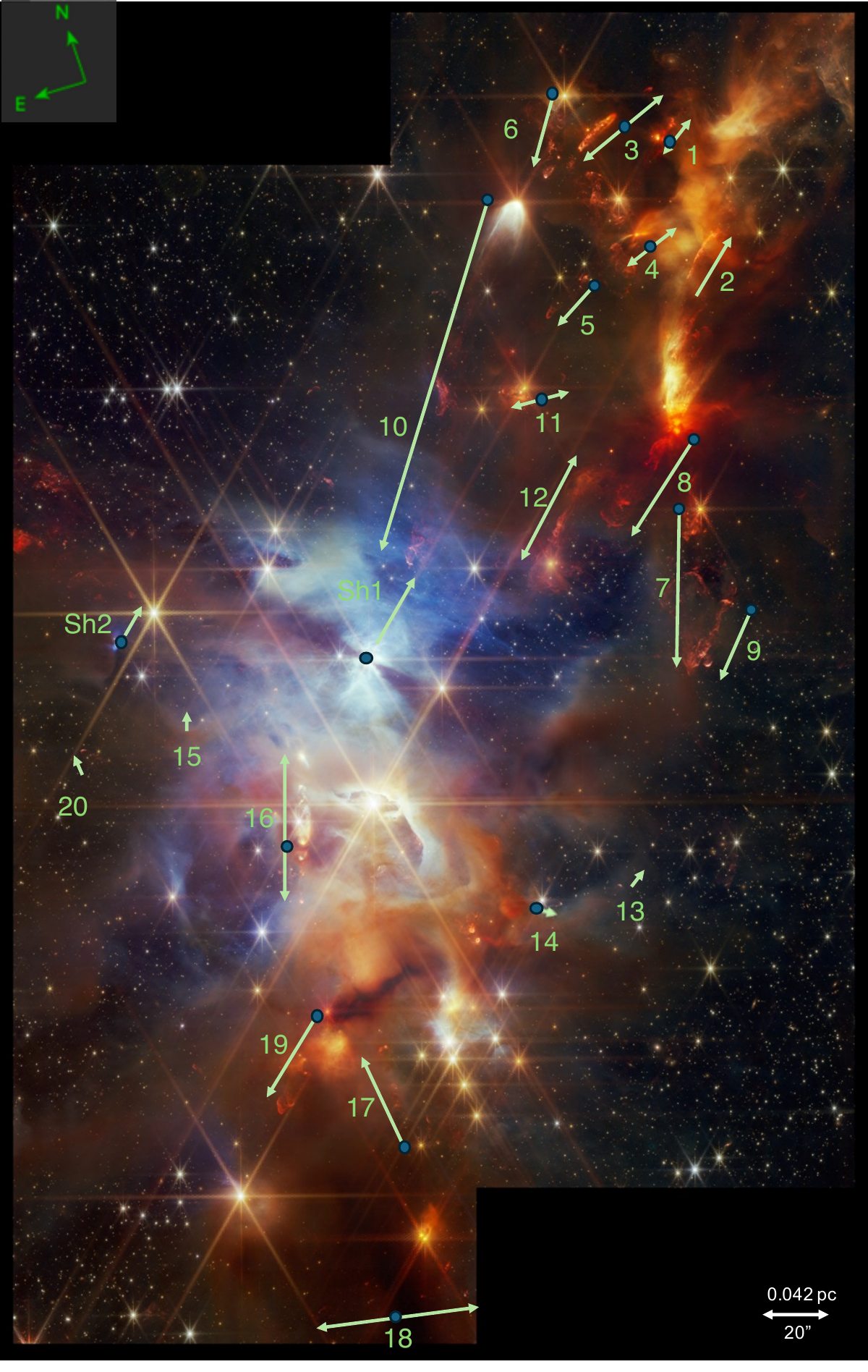}
    \caption{The central location of each outflow (green arrows) and suggested driving sources (blue stars) indicated on a NIRCam-color image (F140M - blue, F210M - green, F360M - orange, F480M - red). The arrow and source locations are offset from the outflow for clarity - refer to the coordinates in the catalogue for accurate outflow coordinates. This combined image is centered at approximately 18:29:55.8 +01:14:34. Image processing credit: Alyssa Pagan.}
    \label{fig:outflowloc}
\end{figure*}

The spin axes of very young stars may be efficiently traced by their outflows. Indeed, the emergence of energetic protostellar outflows is a ubiquitous signature of early star formation \citep{Frank14}. Collimated jets launching from the innermost regions of low-mass young stars impact surrounding molecular cloud material and can create striking structures of shocked ionized, atomic, and molecular gas \citep{reipurth01,bally16}. Since the jets are likely accelerated and collimated by a rapidly rotating poloidal magnetic field in the inner star-disk system, they emerge along the stellar rotation axis and thus trace the angular momentum vector of the star itself \citep{Kwan88,Ouyed97,Banerjee06}. 

Jet material ejected from protostellar systems may contain sufficient momentum to reach distances comparable to the entire cloud, giving rise to spectacular ``parsec-scale'' outflows \citep{Eisloffel97, reipurth97}. As some protostellar outflows traverse molecular cloud core scales ($\sim$1-2\,pc) in less than the cloud life time, they provide an important feedback mechanism that may act to limit the ability of a cloud to form new stars \citep{Hansen12,Plunkett15}. Indeed, molecular clouds are known to form stars at a relatively low conversion efficiency \citep{Evans09,Federrath12}.

Previous searches for correlated protostellar spin axis alignments have had mixed results. For instance, the UWISH2 survey \citep{froebrich16} of Cassiopeia/Auriga and Cygnus X \citep{makin18}
identified a large number of protostellar outflows and found uncorrelated outflow position angles (PAs) on $\gtrsim$10 pc scales.  \citet{baug20} found no alignment in protoclusters in H II regions using ALMA. More recently, using JWST-NIRCam data, \citet{reiter22} also found random orientations of protostellar outflows in NGC 3324 over a field almost 5 pc wide. \citet{Healy23} found no unambiguous signatures of spin alignment in 15 clusters aged 10–100 Myr, using a cluster-wide spatial analysis. \citet{Hull13} did not find evidence for alignment of the magnetic field and outflow axes in protostars. However, \citet{xu22} found that outflow orientations in nearby low mass star forming regions are significantly aligned with dust polarization vectors at 335 GHz measured by {\it Planck} on size scales $>$ 0.5 pc. Further, the individual outflows are well-aligned with their immediate neighbors on these scales. As predicted by models \citep{Misugi23}, \citet{kong19} found evidence of alignment in CO outflows perpendicular to the parent filament. Thus, there is some prior evidence for coherence on core (or filament) size scales that is not found on molecular cloud scales. However, \cite{Hull17} studied a wide range of scales in a single Serpens protostar and did not find that the protostellar structure was aligned with a strong magnetic field.

Statistically complete, wide-field observations of the youngest outflows are challenging because of the high dust extinction in the centers of protostellar cores ($A_V>>10$) and the relatively small fields of view of millimeter interferometers. Thus, while many shock tracers are found in the optical spectrum, these are not visible during the earliest stages of star formation. Conversely, infrared tracers (particularly rotational molecular hydrogen lines like H$_2$ S(9) at 4.8 $\mu$m) are much more accessible, in particular to the high resolution and sensitivity of the James Webb Space Telescope \citep[JWST;][]{Gardner23}. Serpens Main is one of the densest sections of the larger Aquila Rift, consisting of two regions of young stars \citep{eiroa08, Duarte10, herczeg19, pokhrel23},including some of the densest young stellar associations within 500 pc \citep{Pontoppidan04},  with an estimated age of 10$^5$ yr \citep{Harvey07}. Class 0/I sources are found primarily in the subcluster/central regions of both the NW and SE regions while Class II/III sources are spread out across the region \citep{winston07,lee14}. The Serpens filament is known to display a large coherent magnetic field, possibly related to its formation \citep{Kusune19}, making this region a good candidate for connecting alignments of young stars to filamentary structure.  However, previous wide-field imaging of CO outflows in Serpens used single-dish data at too low spatial resolution ($\sim$15\arcsec) to obtain reliable statistics of outflow alignment \citep{Graves10}. In this paper, we present a JWST imaging survey of protostellar outflows in the Serpens Main cluster, and show that the orientations of the outflows are highly non-random, and perpendicular to the magnetic field lines of the Serpens filament. In Section \ref{sec:data} we describe the NIRCam and ancillary observations. Section \ref{sec:analysis} describes the analysis approach and the resulting outflow statistics. Finally, we interpret our findings in Section \ref{sec:discussion}, and conclude with potential implications for the Serpens filament, and other star forming regions.

\section{Observations} 
\label{sec:data}

\subsection{NIRCam image}
We observed the Serpens Main field with the Near-Infrared Camera \citep[NIRCam;][]{Rieke23} on JWST as a pre-image preparing for a Near-Infrared Spectrometer \citep[NIRSpec;][]{Jakobsen22} survey of ices \citep[PID 1611;][]{Pontoppidan21}. We used four medium-band filters, spanning 1.4 to 4.8\,$\mu$m, targeting stellar molecular bands, as well as the 2.12\,$\mu$m rovibrational H$_2$ and 4.69\,$\mu$m rotational H$_2$ S(9) line. The dithering strategy used for the JWST Early Release Observations \citep{Pontoppidan22} were used to optimize the uniformity of the depth over as large a fraction of the field as possible, and to minimize 1/F noise, cosmic rays and bad pixels. Specifically, the image is constructed as a 2$\times$1 mosaic with rows offset by 20\% and with a combined area of approximately 6.6 $\times$ 4.3 arcmin. The maximum total depth in the field is 1800\,s per filter, distributed on 12 dithers and 7 groups using the BRIGHT2 readout pattern. The images were obtained in two visits on 2023 26 Apr and 2023 12 May. We reduced the data using the JWST calibration pipeline \citep{Bushouse23}. However, given the lack of high-quality Gaia astrometric reference stars, we processed the data in two steps. The first step processed the F360M filter with the {\tt tweakreg} step switched off. We then used the {\tt photutils} package to detect point sources and create an astrometric reference catalog. The remaining three filters were then reduced aligning to the F360M catalog to obtain a high-quality relative registration of the image. The absolute frame was then registered with the same offset to a new frame manually adjusted to a combination of Gaia and 2MASS stars. The images were processed with version 11.16.21 of the calibration pipeline and context {\tt jwst\_1084.pmap} of the Calibration Reference Database System (CRDS). {The spatial resolution of NIRCam at 4 $\mu$m is 0$\farcs$13, or about 0$\farcs$16 at F480M.} The properties of the filters are summarized in Table \ref{tab:obs}.

\begin{deluxetable}{lcc}[ht!]
\tablecaption{NIRCam filter summary.
\label{tab:obs}}
\tablehead{
\colhead{Filter} & \colhead{Wavelength} & \colhead{Tracers} \\
\colhead{} & \colhead{$\mu$m} & \colhead{}  
}
\startdata
F140M &1.3-1.5& Reflection nebulosity \\
F210M &2.0-2.2& H$_2$ $v=1-0$ S(1)\\
F360M &3.4-3.8& H$_2$ $v=0-0$ S(14)-S(18) \\
F480M &4.66-5.0& $v=0-0$ H$_2$ S(9) \\
      &        & CO $v=1-0$ P(1)-P(32)   \\
            &        & CO $v=2-1$ P(4)-P(25)   \\
      &        & [Fe II] a4F7/2 - a6D7/2
\enddata
\end{deluxetable}

\subsection{Outflow tracers in NIRCam bandpasses}

Protostellar outflows are generally best detected with NIRCam in the F480M bandpass. This bandpass contains the 4.66\,$\mu$m H$_2$ S(9) line, the 4.89 \,$\mu$m [Fe II]  line, and 54 CO fundamental P-branch lines, known to be strong in protostellar outflows \citep{ray23,federman24,rubinstein24}. Further, this longest wavelength is the least affected by extinction, with optical depths a factor 2.5 lower at 4.8\,$\mu$m compared to 2.1\,$\mu$m \citep{Pontoppidan24}. We consequently use the F480M image to identify candidate outflows by their morphological appearance and to identify knot and bow shock substructures within each outflow (see Figure \ref{fig:outflowloc} for an overview). 

\begin{figure*}[ht!]
    \centering
    \includegraphics[width=0.49\textwidth]{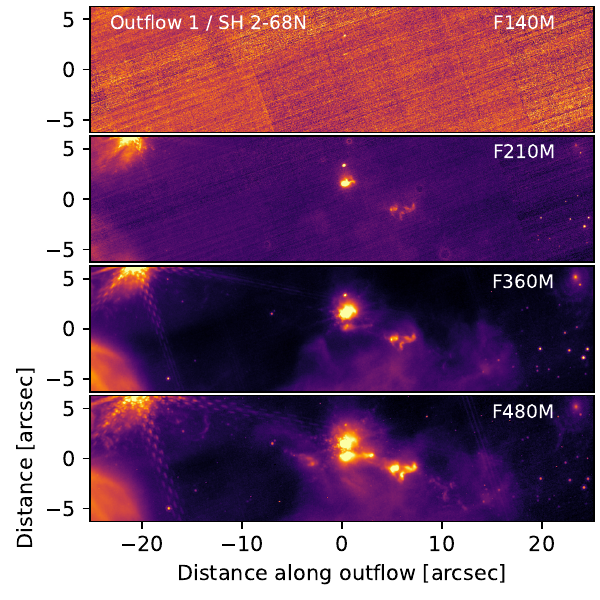}
    \includegraphics[width=0.49\textwidth]{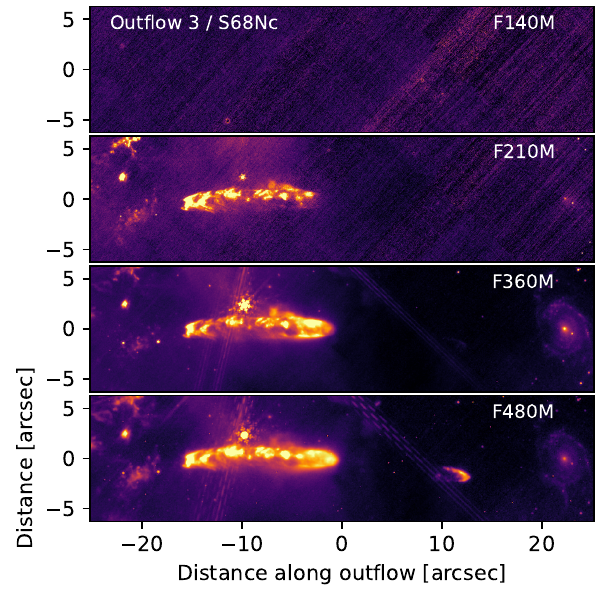}
    \includegraphics[width=0.49\textwidth]{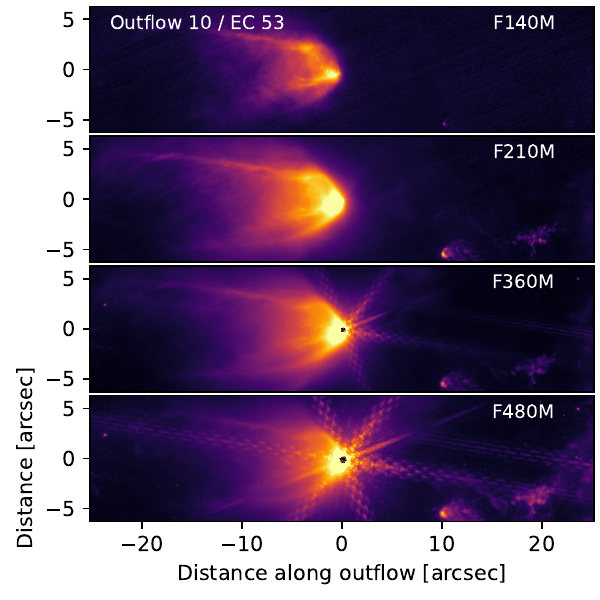}
    \includegraphics[width=0.49\textwidth]{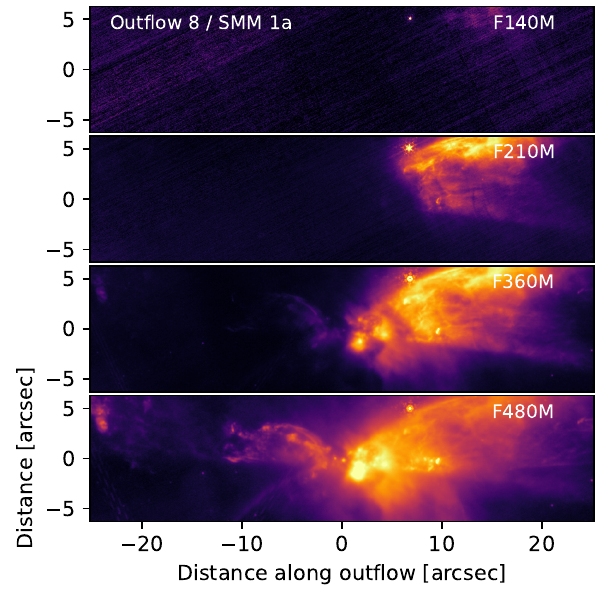}
    \caption{Bandpass comparison of four prominent outflows. The outflows have been rotated with the position angle in Table \ref{tab:pa} to align them with the horizontal axis. At a distance of 430\,pc, the  extent of each image along the x-axis corresponds to 0.1\,pc.}
    \label{fig:filter_comp}
\end{figure*}

We use the F360M band to assist in identifying outflow parameters, but as the line emission is dominated by the weaker rotational H$_2$ S(14)--S(18) lines \citep{ray23}, this band mainly confirms the presence of an outflow (see Figure \ref{fig:filter_comp} for a comparison). The detailed similarity of outflow candidates in the F360M and F480M bands supports that H$_2$ S(9) is the most likely dominant source of emission in F480M. 

The extended emission in the F210M band is likely dominated by H$_2$ v$=$1-0 rovibrational emission from S(0) to S(4), with a contribution from reflection nebulosity. However, this band may also contain Br $\gamma$ emission, which could come from irradiated cloud edges or dissociative shocks and is not easily separated from molecular emission. On the other hand, based on the similarity of the emission in the F480M band, we assume that the H$_2$ dominates both filters. Rovibrational H$_2$ lines are excited under different conditions than the rotational H$_2$ S(9) line, and suffer from greater extinction. Consequently, only a subset of outflows appear clearly in both F210M and F480M (Figure \ref{fig:filter_comp}). 15 of 20 outflows are observed in F210M, although 5 of those 15 are only partially detected compared with the full F480M morphology.

Finally, the F140M band is dominated by reflection nebulosity, with prominent illuminating sources such as EC 82 (the Great Disk Shadow; \citealt{pontoppidan20}) and EC 90 lighting up the SE region. We use the two disk shadows seen in the reflection nebulosity to augment our sample of protostars with measured position angles (see Section \ref{subsec:shadow}), and are identified as Sh1 and Sh2 respectively in Figure \ref{fig:outflowloc}. We summarize the tracers in each filter in Table \ref{tab:obs}. Most of the north-eastern core is not visible in F140M due to extinction. 

\subsection{Polarization Maps} 

We use archival SOFIA-HAWC+ data to sample the orientation of the cloud-scale magnetic field in Serpens Main. The Serpens Main region was observed in Band D ($\sim$ 154 $\mu$m) and Band E ($\sim$ 214 $\mu$m) with HAWC+ on flight F621, on 10 Oct 2019, as a part of the SOFIA Cycle 7 program 0130 (PI: L. Fanciullo). Serpens Main was observed on this flight using the On-The-Fly Mapping (OTFMAP) scan mode of HAWC+. A Lissajous scan pattern with scan angle of -30 $\deg$, scan amplitude of 220 arcsec with a slew rate of 200 arcsec/s was used to obtain this data. Multiple pointings (4 in Band D and 7 in Band E) were used to cover an area of 13 $\times$ 13 arcmin$^2$ of the Serpens Main star-forming region with a total integration time of 1952 and 3555 sec in Bands D and E, respectively. This resulted in higher signal-to-noise ratio (SNR) in Band E compared with Band D. Therefore we used Band E for our best sampled dataset to investigate the B-field orientation around our sample.

The HAWC+ Band E data was re-processed using the SOFIA Data Reduction software, SOFIA\_Redux Version 1.3.3 (HAWC+ DRP Version 3.2.0). The resulting level 4 mosaic of HAWC+ Band E polarization maps have a pixel size of 4$\farcs$55 and effective beam size of 18$\farcs$2. The final level 4 data product includes Stokes parameters I, Q and U, the polarization fraction P, the polarization angle $\theta$ and their uncertainties. Since the thermal emission from interstellar dust grains is preferentially polarized perpendicular to the  magnetic field, the direction of the magnetic field in the plane of the sky can be obtained by adding $\pi/2$ to the polarization angle $\theta$ and is included in the level 4 mosaic  (\citealt{hoang14,andersson15} and references therein). For a detailed calculation of each of these quantities we refer the readers to the HAWC+ DRP User's Manual and \citet{gordon18}.

To ensure the highest quality polarization measurements and exclude low SNR pixels, we masked our Band E array, including only pixels with SNR $\ge$ 150 in total intensity (Stokes I), $< 50\%$ in percent polarization, and a SNR of $>$ 3 in polarization fraction. We measured the average polarization angle in a half beam (9$\farcs$1 radius circle) around each of our targets and include it in Table \ref{tab:catalog}.

\section{Analysis}
\label{sec:analysis}

\subsection{The Serpens Main cluster}
Figure \ref{fig:outflowloc} shows a color-composite of the NIRCam image. The NW and SE regions together form a flow axis that constitutes the Serpens Main region; considerably off the south edge of the mosaic is Serpens South \citep{Gutermuth08}. It is clear that the NW and SE regions contain the densest and most opaque material in this region.

\subsection{Identifying outflows}

\begin{figure*}
    \centering
    \includegraphics[width=0.99\textwidth]{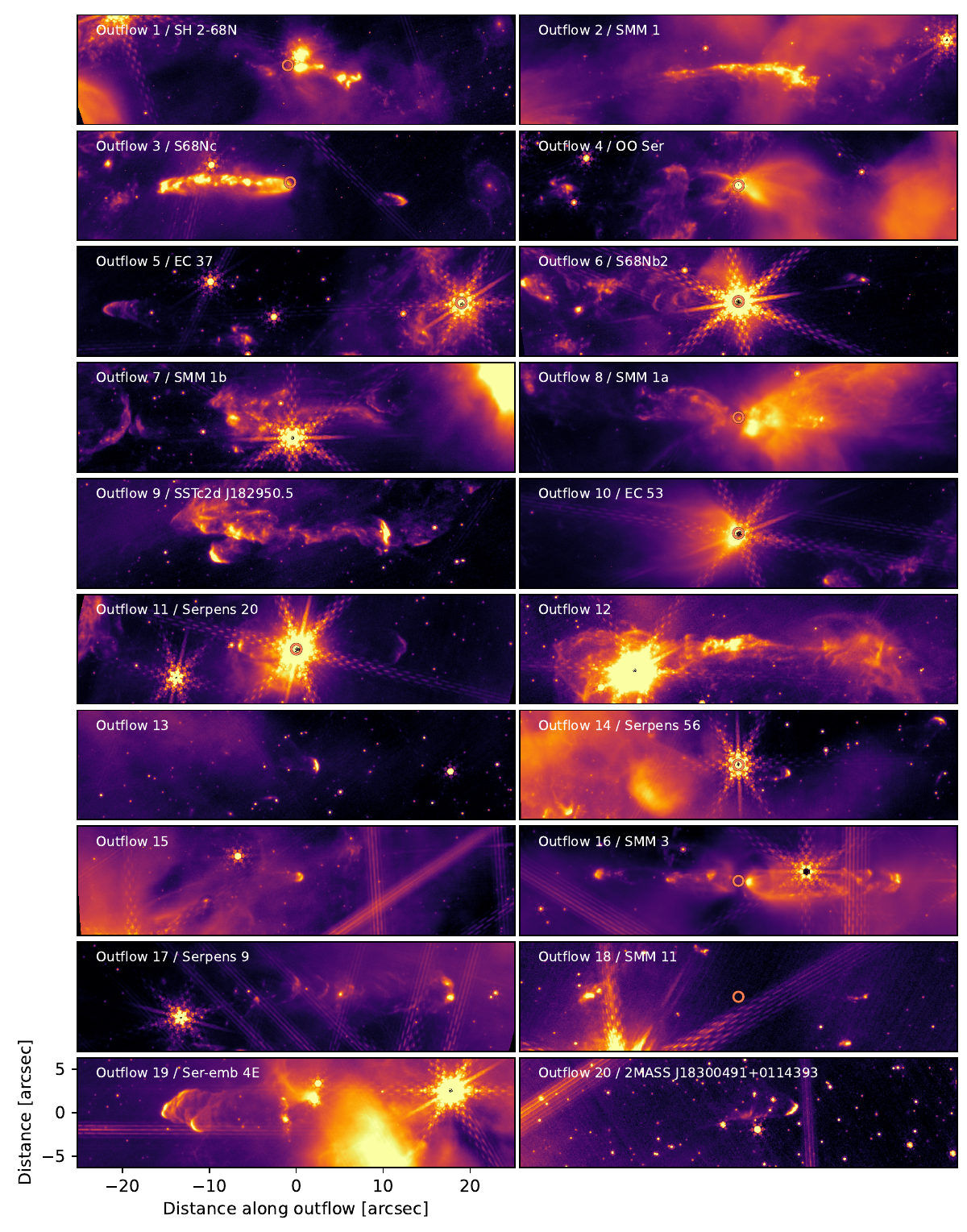}
    \caption{A gallery of the F480M images of each outflow. The scale of each image is identical, and outflows have been rotated by the PA provided in Table \ref{tab:pa}. {\bf Orange} circles indicate the position of the driving source candidate, when known. The images are scaled using an arcsinh function to emphasize faint, extended emission.}
    \label{fig:outflowgallery}
\end{figure*}

It is visually apparent from Figure \ref{fig:outflowloc} that most outflows in the region appear to be aligned in position angle. However, to quantify the alignment, we identify outflows in the NIRCam images based on a hierarchy of criteria. Using the F480M image, which includes the strongest and least extinguished outflow lines, we visually searched for extended structure with a ``bow shock'' type morphology, defined as a $\sim$180 degree `C' shaped arc. Since the bow shocks are directional, we tracked each backward until locating either: 1) another bow shock with similar orientation, 2) a series of compact knots indicative of a jet along the same orientation, or 3) a continuum source that could plausibly be driving the outflow. Any system meeting this criteria is collectively considered an outflow candidate (lowest confidence class C). For each candidate, the F360M and F210M images were inspected for counterparts to the bow shocks seen in the F480M image. If the outflow is recovered in at least one of the F210 or F360M filters (but not F140M, which does not typically reveal outflows due to extinction and lack of H$_2$ lines), the outflow candidate is given confidence class B. Finally, if 1) a driving source can plausibly be identified, or 2) another bow shock oriented in the opposite direction, and along the outflow axis is detected, the outflow candidate is given the highest confidence class A. Although the catalog includes outflow candidates from all confidence classes, only those with confidence A are included in our statistics in the following analysis. The location of each outflow is shown in Figure \ref{fig:outflowloc}, an aligned gallery is shown in Figure \ref{fig:outflowgallery}, and the catalog itself is presented in Table \ref{tab:catalog}. 

\begin{deluxetable*}{llcrccrccccl}[ht!]
\tablecaption{Average position angle and uncertainty, and likely driving source for each outflow in this work. RA/Dec are given for the central/driving source coordinates.  Pol. is the dust polarization angle as measured in the HAWC+ Band E (216 $\mu$m archival data (see text).
\label{tab:catalog}}
\tablehead{
\colhead{ID} & \colhead{RA} & \colhead{Dec} & \colhead{PA} & \colhead{Length} & \colhead{Length Ratio\tablenotemark{a}} & \colhead{Pol.\tablenotemark{b}} & \colhead{Conf.} & \colhead{Driving Source Cand.}\\
\colhead{} & \colhead{degree} & \colhead{degree} & \colhead{degree} & \colhead{arcsec} & \colhead{} & \colhead{degree} & \colhead{}
}
\startdata
1  & 277.45017 & 1.27892 & $141.2 \pm 9.3$  & 0.39       & 1.12 & $119.0 \pm  3.5$   & A & SMM 9 (SH 2-68N) \\
2  & 277.45025 & 1.26917 & $129.6 \pm 3.5$  & 0.59       & 1.11 & $110.5 \pm  3.1$ & B & SMM 1 (S7)\\
3  & 277.45296 & 1.28233 & $112.0 \pm 2.1$  & 0.91       & 1.23 & $118.5 \pm  4.6$ & A & S68Nc$^c$ \\
4  & 277.45471 & 1.27225 & $108.3 \pm 21.4$ & 0.9-2.8    & 1.18 & $110.8 \pm  3.5$ & A & OO Ser \\
5  & 277.45521 & 1.275431 & $115.6 \pm 2.5$  & 1.4        & 1.06 & $109.3 \pm 11.1$ & A & EC37 (V370 Ser)\\
6  & 277.45663 & 1.28506 & $151.6 \pm 2.7$  & 1.3        & --   & $259.2 \pm  5.2$ & A & S68Nb2 \\
7  & 277.45704 & 1.24914 & $158.8 \pm 6.9$  & 0.51-1.7   & --   & $166.0 \pm  4.5$ & A & SMM 1b \\
8  & 277.45742 & 1.25581 & $135.2 \pm 6.2$  & 1.5-5.1    & 1.03 & $131.3 \pm  4.5$ & A & SMM 1a \\
9  & 277.45946 & 1.23919 & $135.6 \pm 9.2$  & 0.89-1.4   & 1.04 & $179.8 \pm  5.3$ & A & SSTc2d J182950.5+01141 \\ 
10 & 277.46321 & 1.27800 & $138.9 \pm 5.2$  & 1.7        & --   & $116.1 \pm  9.8$ & A & EC 53 \\
11 & 277.46742 & 1.26347 & $83.4 \pm 0.6$   & 1.3        & 1.29 & $265.1 \pm  6.7$ & A & Serp 20 \\
12 & 277.46833 & 1.25169 & $132.7 \pm 6.4$  & 1.2        & 1.02 & $106.6 \pm  4.1$ & A & No identification \\
13 & 277.47400 & 1.22158 & $123.2 \pm 15.8$ & 0.59       & 5.56 & $161.7 \pm  6.2$ & B & No identification\\
14 & 277.47996 & 1.22283 & $68.1 \pm 5.0$   & 0.63       & --   & $177.3 \pm  6.2$ & A & Serpens 56 \\
15 & 277.49504 & 1.24622 & $156.3 \pm 11.9$ & 0.33       & --   & $240.1 \pm  6.3$ & B & No identification\\
16 & 277.49642 & 1.23522 & $160.8 \pm 0.7$  & 1.4        & 1.32 & $239.1 \pm  5.1$ & A & SMM 3\\
17 & 277.49646 & 1.21064 & $2.7  \pm 4.8$ & 1.2        & --   & $ 228.9 \pm  3.7$ & A & Serpens 9\\
18 & 277.50167 & 1.19583 & $76.1 \pm 2.0$   & 1.1        & 1.02 & $ 259.0 \pm  6.6$ & B & SMM 11 \\
19 & 277.50296 & 1.21603 & $130.4 \pm 9.1$  & 1.7        & --   & $ 266.9 \pm  4.3$ & C & Ser-emb 4E \\
20 & 277.51067 & 1.24542 & $216.8 \pm 10.1$ & 1.7        & --   & $ 197.3 \pm 13.2$ & A & 2MASS J18300491+0114393 \\
\hline
21 & 277.48688 & 1.24633 & $134.0 \pm 5$  & --        & --   & $ 177.3 \pm  5.1$ & A & [EC92] 82 \\
22 & 277.50621 & 1.25431 & $140.4 \pm 5$  & --        & --   & $ 241.0 \pm 39.5$ & A & Shd 2
\enddata
\tablenotetext{a}{The ratio of the lengths of two outflow lobes. This is only available for bipolar morphologies.}
\tablenotetext{b}{The position angle of the polarization vector.}
\tablenotetext{c}{The driving source position (S68Nc) presented here is the center of the central knot, as indicated in Figures \ref{fig:filter_comp} and \ref{fig:outflowgallery}.}
\end{deluxetable*}

\subsection{Measuring position angles}

\begin{figure}
    \centering
    \includegraphics[width=0.49\textwidth]{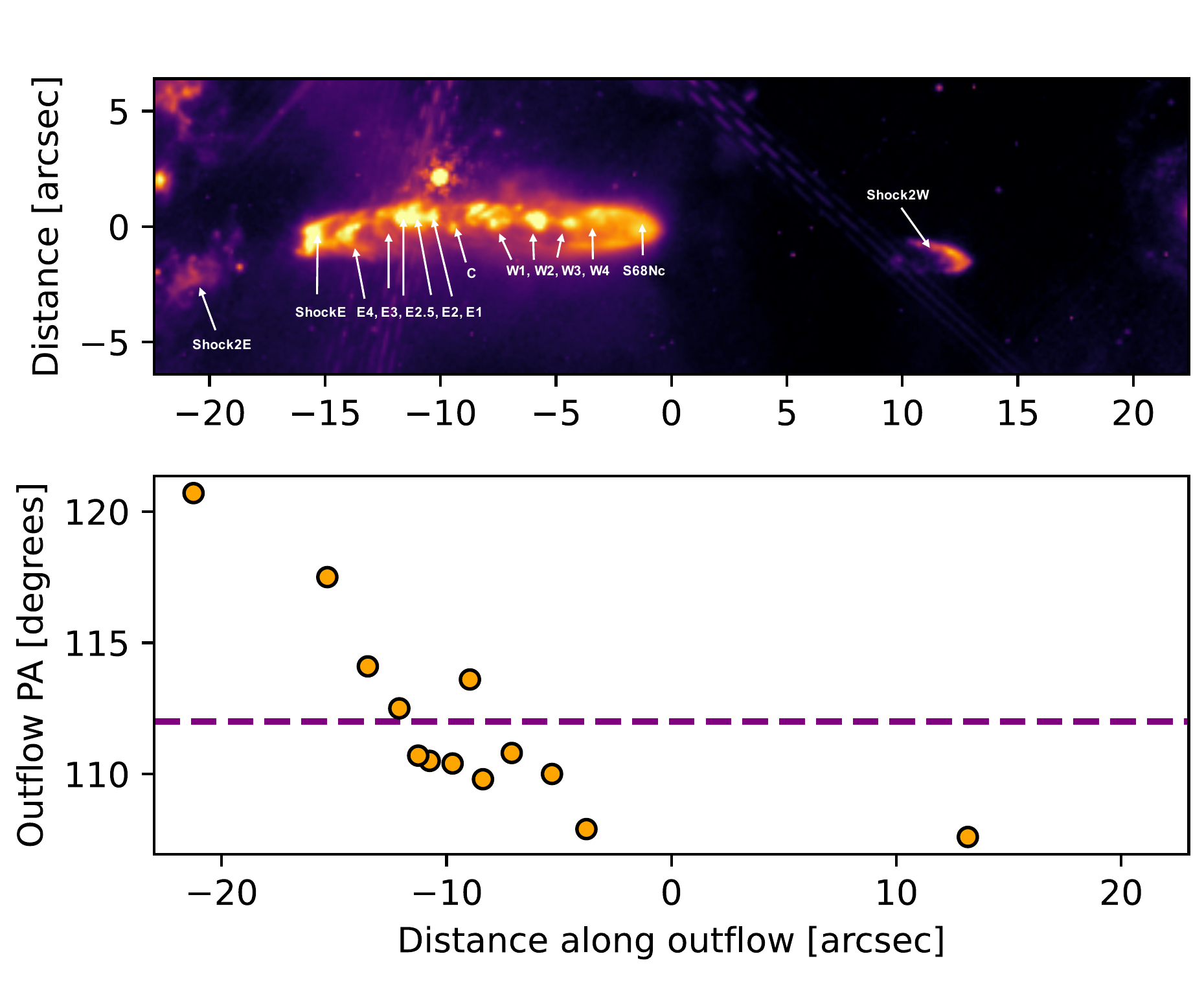}
    \caption{{\bf Top:} Annotated F480M image of the knots composing Outflow 3. {\bf Bottom:} Position angle of each identified knot (relative to the central position of S68Nc) at their respective radius along the outflow axis. The horizontal dashed line is the average position angle.}
    \label{fig:outflow1annot}
\end{figure}

The outflow PAs are measured relative to the candidate driving source, or a central position within the outflow itself if no unambiguous driving source can be identified. For outflows without an obvious driving source, the central position is based on the orientation and position of knots and bow shocks. The central positions are listed in Table \ref{tab:catalog}. The PA is estimated by calculating a separate PA from the driving source to each identified knot in the flow (see Figure \ref{fig:outflow1annot}). These are then averaged to produce a single value. To estimate the uncertainty in PA, we take the width of the outermost bow shock edge and calculate the range of allowable angles relative to the central/driving source. For outflows with clearly defined morphologies, this uncertainty varies between 1 and 10\degree, but is as high as $\sim$ 20\degree\ for nebulous, wide angle, or overlapping flows. The longer an outflow is, or the narrower the morphology appears, the better constrained the PA becomes. Thus, outflows or tightly collimated jet-like structures with clear driving sources have the lowest uncertainty. 

An example of the identified knot structures used for the PA determination is provided for one outflow in Figure \ref{fig:outflow1annot} and Table \ref{tab:pa}. In this case, some of the change of PA knot-to-knot appears systematic, perhaps due to precession, suggesting that our PA uncertainty estimate is slightly conservative.

\begin{deluxetable}{cllcc}[ht!]
\tablecaption{Location, PA, and distance from center (bright source) position of Outflow 3. 
\label{tab:pa}}
\tablehead{
\colhead{Knot} & \colhead{RA} & \colhead{Dec} & \colhead{PA} & \colhead{Dist} \\
\colhead{} & \colhead{degree} & \colhead{degree} & \colhead{degree} & \colhead{arcsec}
}
\startdata
        Shock2W & 277.44954 & 1.28342 & -72.4 (107.6) & 13.18   \\
                \hline
        S68Nc & 277.45296 & 1.28233 & -- & 0  \\
                        \hline
        W4 & 277.45392 & 1.28203 & 107.9 & 3.78   \\ 
        W3 & 277.45433 & 1.28183 & 110.0  & 5.31   \\
        W2 & 277.45479 & 1.28164 & 110.8 & 7.10  \\
        W1 & 277.45513 & 1.28156 & 109.8 & 8.38   \\
        C & 277.45521 & 1.28136 & 113.6 & 8.96  \\
        E1 & 277.45546 & 1.28142 & 110.4 & 9.73 \\
        E2 & 277.45571 & 1.28131 & 110.5 & 10.75  \\
        E2.5 & 277.45583 & 1.28125 & 110.7 & 11.26\\
        E3 & 277.45600 & 1.28108 & 112.5 & 12.10  \\
        E4 & 277.45633 & 1.28083 & 114.1 & 13.50  \\
        ShockE & 277.45675 & 1.28058 & 117.5 & 15.30 \\
        Shock2E & 277.45813 & 1.27928 & 120.7 & 21.25 \\
\enddata
\end{deluxetable}

\subsection{Position angles for edge-on disks}
\label{subsec:shadow}

There are two edge-on disks in the field that supplement the source position angles indicated by the outflows: EC 82 and ``Shadow Jr.'' (or ``Shd 2'', as referred to in this work; see Figure \ref{fig:outflowloc}). The disk around the intermediate-mass young star EC 82 casts a large shadow on surrounding reflection nebulosity, giving rise to the so-called ``Great Serpens Disk Shadow'', first analyzed using data from the Hubble Space Telescope \citep{pontoppidan20}. The expansive shadow is most noticeable in the F140M image.  Because the disk position angle is well-established, it represents a complementary star for which the rotation axis is likely known, assuming it is traced by its disk. Additionally, a second, much smaller, disk shadow, noted in \citet{pontoppidan20}, is also visible east of EC 82. The orientation of this second disk shadow is similar to that of EC 82. Although we do not clearly detect jets/outflows around these two sources in the F480M data, they cannot be ruled out. Both angles are provided in Table \ref{tab:obs}, rotated by 90$\degree$ to match the outflow axes for the rest of the sample, assuming these are perpendicular to the disk. 

\subsection{Outflow Dimensions}

The width of each outflow is measured from the terminus or shockfront knots of emission, where the cavity should be at its widest, perpendicular to the outflow position angle until clearly defined walls of the outflow cavity can no longer be easily distinguished from background nebulosity; for an illustration of these parameters, see the top part of Figure \ref{fig:outflow1annot}.  In this example, the Shock2W position represents the point of the bow shock. We measure the full width of the bow shock by visual identification of where each side is detected above the background. We perform a similar estimation for each outflow knot. We may observe a weak but positive correlation between outflow length and width, but in general conclude that these parameters are not predictive of each other.

The length of the outflows with bipolar morphology varies considerably, from $\sim 9-65 \arcsec$. At a typical 430 pc distance to Serpens Main \citep{herczeg19}, assuming a shock speed of 100 km/s \citep{reiter22} we find that the dynamic age of the outflows ranges from 200 - 1400 yr, considerably younger than many of the outflows in the NGC 3324 study, which generally found kinematic ages of 1000--10000\,yr.

\section{Discussion}
\label{sec:discussion}

\subsection{Outflow Density}

The surface density of young stars of all classes in Serpens Main has been estimated at 79 YSO per pc$^2$ \citep[scaled to the correct distance to Serpens;][]{Harvey07}. The 20 outflows we are identify are contained in a region measuring approximately 0.6 pc $\times$ 0.5 pc, or about 66 outflows per pc$^2$. This is considerably higher density of flows than in other star forming regions observed with NIRCam. Carina (NGC 3324) included about 31 identified outflows in a roughly 3 pc $\times$ 2 pc region \citep{reiter22}, or about 5 outflows per pc$^2$, more than a factor of ten lower than in Serpens. This may be attributable to a number of effects, including differences in resolution (NGC 3324 is eight times the distance of Serpens), age of the clusters, and prevalence of nearby massive stars. In NGC 1333, a comparably-sized low mass cluster, \citet{Knee00} identify 10 outflows using rotational CO mapping of a 0.65 pc$^2$ region, corresponding to a density of 15 outflows per pc$^2$). This suggests that NIRCam is a powerful instrument for surveying protostellar outflows in nearby star-forming regions.  

\subsection{Position Angle alignment}
\begin{figure}
    \centering
    \includegraphics[width=0.49\textwidth]{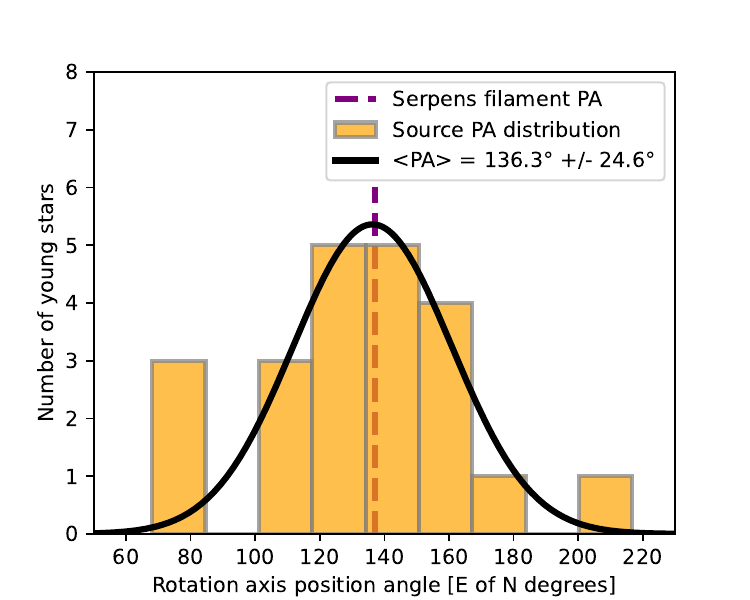}
    \caption{Distribution of measured average position angles for all 22 sources, clustering around the filament PA = 139\degree. The black curve is a Gaussian fit to the distribution with parameters (mean and standard deviation) given in the legend.}
    \label{fig:pahist}
\end{figure}

The measured average position angle for each of the 20 outflows and the 2 disks (assuming the outflow axis is 90\degree\ to the disk axis) is tabulated in Table \ref{tab:catalog} and the distribution of these 22 angles is shown in Figure \ref{fig:pahist}. Considering only the 15 high confidence outflows (class A), at least 8 are aligned to within $\pm$10\degree. The two disk shadows in this region have angular momentum axes that are aligned with the outflows adding to the total of 10 of 17 high confidence orientations falling within a $\pm$10\degree\ span. Further, 14 of 17 objects have PAs falling within a $\pm$30\degree\ span.

We used a simple Monte Carlo analysis to test the null hypothesis that the catalogued outflow orientations are randomly distributed. For all of the calculations here, we assumed a uniform distribution of outflow PAs between 0 and 180\degree.  To determine the likelihood of the PA distribution arising randomly from a uniform distribution, we used the {\tt numpy} random number generator to produce 10$^5$ instances and note the number of occurrences with at least the observed PA clustering. The odds of 10 of 17 uniformly distributed sources falling in a single 20\degree\ bin is $\sim$ 0.002\%, and the odds of 14 of 17 high confidence sources being aligned in a 60\degree\ bin is only slightly higher at about 0.005\%.

\begin{figure}
    \centering
    \includegraphics[width=0.49\textwidth]{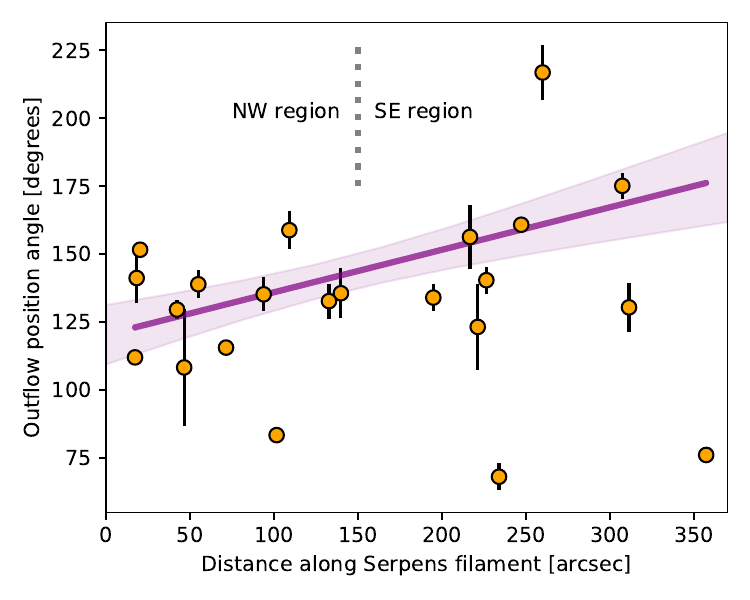}
    \caption{PA as a function of position along the filament. The PAs of the NW region are more correlated than the SW region. The line is the best linear fit after removing three outliers with the highest and lowest measured PA. The shaded region shows the 99\% confidence level of the fit.}
    \label{fig:loc_corr}
\end{figure}

Figure \ref{fig:loc_corr} shows the distribution of outflow position angles as a function of driving source position along the axis of the Serpens filament. The axis is estimated to be at PA of 139\degree, along the line connecting the centers of mass from the SW to the NE regions from the FIR imagery of the Serpens region / Aquila Rift \citep{Gong21}. This parameter is used as a measure of location along the filament; north-west to south-east. There is a strong correlation with the north-western part of outflows clustering in position angle around a mean of 136\degree.

\subsubsection{Are the outflows at similar inclination angles?}

Outflows 1-4. 7-9, 11 and 12 are all bipolar, with their lobe length ratios between 1.02 and 1.29 (ie. 2 - 29\% deviation from perfect symmetry). Although symmetrical lobes are not a direct indicator of an edge-on inclination system, because of the local extinction or distortion through interaction with the local cloud material, it is likely that strongly inclined outflows would not show such symmetry across the sample. For example, \citet{Habel21} consider this criteria in identifying bipolar outflows with more edge-on systems. The close symmetry is at least consistent with relatively edge-on, and therefore relatively similar inclination angles. Considering this inclination constraint along with the tight clustering of position angles, this supports the idea that these outflows are similarly oriented in 3-dimensional space. However, as many of these outflows extend considerably beyond common protostellar envelope scales, or have asymmetric structures close to the driving source, bipolar symmetry at large distances is suggestive rather than conclusive.

\subsection{Outflow orientation vs. dust polarization vectors}

To compare the filament and individual outflow orientations with the larger scale magnetic field, we compared our results with archival datasets. First we compared Figure 2 from \citet{kwon22} - their map of the inferred magnetic field vectors - with our NIRCam mosaic. It was immediately apparent that the magnetic field lines were roughly perpendicular to the outflow direction in the NW region, but are less organized and systematic in the rest of the field, except along the identified filaments from the \citet{kwon22} analysis.

\begin{figure*}[ht!]
    \centering
    \includegraphics[width=\textwidth]{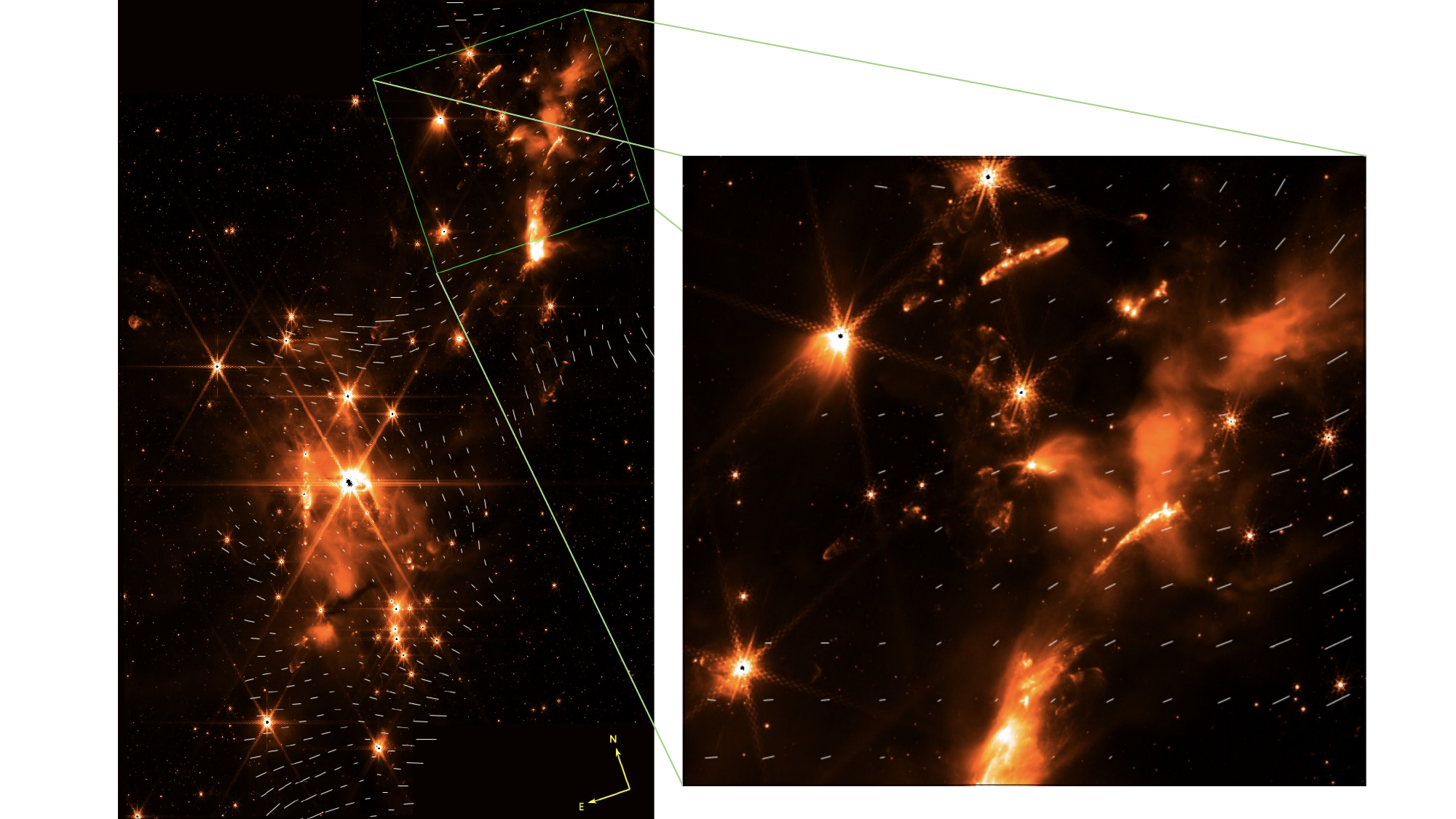}
    \caption{{\bf Left:} Overlay of HAWC+ Band E polarization vectors (white arrows) on the full 6.6 $\times$ 4.3 arcmin NIRCam F480M image from this work. The SOFIA vectors match well with SCUBA maps from \citet{kwon22}. {\bf Right:} Zoom on the NW region filament, where the most aligned outflows are located. This inset region spans approximately 1.7 $\times$ 1.9 arcmin (N-S $\times$ E-W extent, respectively).}
    \label{fig:hawc}
\end{figure*}

To improve the resolution and better resolve individual driving sources/cores, we re-reduced and interpolated the HAWC+ Band E dust polarization vectors to the positions of each of the 22 center positions, shown in Table \ref{tab:pa}, and displayed them in Figure \ref{fig:hawc} overlaid on the F480M NIRCam image.  For display purposes, we scaled the lengths of the polarization vectors for easier visual comparison with outflow orientations. It is apparent that the Band E vectors closely track most of the outflows, and a comparison of the position angles in Table \ref{tab:catalog} confirms this. Of the 12 outflows in the NW region (see the rightmost panel of Fig \ref{fig:hawc} for a zoomed-in view), all but 2 are within 25 degrees of alignment with their respective magnetic field polarization vectors. One of those 2 (outflow 9) has no identified driving source.  The other, outflow 6, is the only significant outlier in this region. The alignment with the magnetic field in the SE region is less correlated. Of outflows 13-20, only 2 (outflows 18 and 20) are closely aligned with the nearby polarization vectors. The polarization vectors do not align with the two disk shadows either. Conversely, the polarization fraction, indicated by the vector length, is larger in the SE region, and around the disk shadows, than it is in the NW region.

\subsection{Comparison to outflow alignments in other regions}

Outflow surveys in other star-forming regions often find no preferred outflow orientation, but typically on much larger scales than Serpens Main \cite[5-10 pc; ][]{stephens17,baug20,reiter22}. The sensitivity and spatial resolution of NIRCam to detect a statistically significant number of outflows on scales smaller than $\sim$1\,pc may explain, in part, why we detect the alignment in Serpens Main. Indeed, there is existing evidence of relative alignment between outflow axes on such scales for the youngest clusters in filaments \citep[e.g.,][]{davis07,kong19}. Thus, we suggest that our NIRCam image indicates that alignment has a coherence scale of $\lesssim 1$\,pc, and that alignment is rapidly degraded with time due to precession and binary interactions.  Misalignment processes are predicted to occur on timescales of 10$^5$ - 10$^6$ yr \citep{Lai14}. If these effects randomized spins on timescales much shorter than that, the observed alignment would not be possible. While \citet{Misugi23} predicts that the core rotation axes (not necessarily individual outflows) are perpendicular to the filament, and \citet{kong19} find an example of this, this is inconsistent with the dominant orientation of the Serpens outflows, which instead appear to be aligned with the filament axis (Figure \ref{fig:pahist}). However, the large-scale orientation of the dynamical filament in the Serpens Main region may be different than the simplified axis defined by the vector between the NW to SE clusters. The filament seen in the extinction map in \cite{Fiorellino21} (their Figure 15) presents an arc, rather than a linear structure, suggesting a more complex arrangement in which the orientation of the filament potentially changed since the initial fragmentation of the cluster. Thus, while we note the discrepancy in the protostellar alignment with the apparent filament orientation compared to the theoretical expectation (parallel rather than perpendicular), this is not necessarily strong evidence against the theoretical prediction. Further dynamical modeling is required to explain the apparent parallel alignment of the Serpens outflow axes with the elongation of the local Serpens cloud.

How are alignments related to the magnetic field? Recently, \citet{xu22} showed that outflow orientations are not random compared to the large-scale magnetic field. We see close alignment in magnetic field orientation and the outflows in the NW region presented here, but not in the SE region. These alignments suggest that the large-scale magnetic fields that help funnel material onto filaments also determine the initial orientation of the outflow axes. Observations suggest less outflow alignment over time as stellar feedback disrupts the magnetic field alignment and anisotropic accretion alters the outflow axes of the embedded protostars. An alternate, larger scale effect could be cloud-cloud collisions. \citet{Duarte10} also identified the NE region as containing more uniform conditions for young stars, but they argued instead the SE region was ``perturbed'' by a cloud-cloud collision in progress, while the NW region was ``homogeneous''.  We argue here that the alignment of spin axes is further evidence for a lack of perturbation of the NW clump. Alignments are more pronounced in young regions \citep[e.g.,][]{kong19} while there is less evidence for a preferential outflow direction in older regions and those significantly affected by stellar feedback \citep[e.g.,][]{feddersen20}. This suggests that outflow alignment may be common in young regions but quickly disrupted. The disruption time likely depends on the strength of the magnetic field and the density of the region. 

Serpens Main is similar to Ophiuchus in age, mass, and average density \citep{Evans09}. However, \citet{xu22} find a larger range of CO outflow position angles in Ophiucus than we find in Serpens Main. Millimeter CO emission tends to trace less collimated outflow components than the infrared emission presented in this paper. Using the same outflow tracers would provide a more direct comparison of the outflow orientation of these regions.

Weaker fields may also lead to less outflow alignment in a given region. 
\citet{xu22} propose that weaker field strengths may contribute to the lack of outflow alignment in Perseus \citep{stephens17}. If true, this predicts a stronger magnetic field in Serpens Main \citep[$\approx 60 - 300\mu$G;][]{kwon22}. However, a more direct comparison of the degree of outflow alignment with the local magnetic field is required to test this hypothesis.  Nevertheless, our results are consistent with several other studies that find a higher degree of outflow alignment in the youngest, darkest regions of the cloud \citep[e.g.][]{davis07,makin18}.  

\section{Summary}

We observed the Serpens Main star forming region with JWST-NIRCam, at 1.4, 2.1, 3.6, and 4.8 $\mu$m. In addition to new views of the star forming complex, the images were sensitive to protostellar outflows.

We identified 20 outflows by their bow shock morphology and ancillary data on driving sources, developing a catalog of outflows including knot locations, radii, length, and position angle. 15 of the 20 outflows fall into our highest confidence detection bins, with identified driving sources, most noted in previous surveys. We examined dust polarization images taken by SOFIA/HAWC+ to provide magnetic field alignment and context, considering published ancillary measurements from JCMT-SCUBA, ALMA, and Spitzer-IRAC.

We analyzed the outflows and summarize our results below:

\begin{itemize}
    \item NIRCam/F480M is particularly well-suited to detect outflows because it contains molecular, atomic, and ionic tracers that all emit strongly in protostellar outflows/jets. The result is a mixed morphological catalog with a high detection rate. 
    \item 12 outflows were identified in the northwestern filament/region, while 8 outflows were identified in the southeastern filament/region. Additionally, two prominent disk shadows were confirmed in the central region.
    \item The axes of the 12 outflows in the NW region are inconsistent with random orientations and align with the filament direction from NW to SE. Additionally, the position angle of jets/outflows from the 2 identified disk shadows align with the filament axis. We estimate $<$0.005\% probability of the the observed alignments if sampled from a uniform distribution in position angle. 
    \item The position angles of the outflows align with SOFIA/HAWC+ 214 $\mu$m dust polarization vectors measured locally around each driving source. However, the disk shadows do not align with their local magnetic fields. This broad alignment does not apply in the SE region. Few of the 8 identified outflows in this region align with the filament axis, or with the dust polarization vector.
    \item The density of outflows detected in this catalog ($\sim$ 66 outflows per pc$^2$) is higher than other low mass star forming regions (e.g., NGC 1333), and ten times greater than observed by JWST/NIRCam in Carina (NGC 3324).
\end{itemize}

The alignment of outflows with the filament axis in part of Serpens Main, but not in the rest of the region, is suggestive of an evolutionary process. It appears that star formation proceeded along a magnetically confined filament that set the initial spin for most of protostars. We hypothesize that in the NW region, which may be younger, the alignment is preserved, whereas the spin axes have had time to precess or dissociate through dynamic interactions in the SE region. The disk shadows, which may be more evolved sources, appear to have retained their spin axis relative to the original field lines, but the magnetic field itself has shifted, or the material from the early formation period has notably dispersed (evident by their scattered light emission in F140M) after their development phase.

Above all, this work shows that even a single pair of JWST/NIRCam images in medium bands can provide considerable insight into the history of star-forming regions. We anticipate more detailed studies of star forming filaments with JWST in the future.

\clearpage
\begin{acknowledgments}
The authors would like to thank Nicole Arulanantham, Sylvia Baggett, Neal Evans, Will Fischer, Nicole Karnath, Tom Megeath, Stella Offner, Amanda Pagul, and Adam Rubinstein for helpful discussions and insights. We also thank Alyssa Pagan for producing a beautiful composite NIRCam image of Serpens Main. We thank the JWST Program 1611 team for the use of their proprietary pre-imaging data. We thank Lapo Fanciullo and their team for taking the SOFIA-HAWC+ data, and the SOFIA data pipeline developers for enabling us to quickly reduce the archival data. We thank the anonymous referee for a thorough and considered report that significantly improved the final manuscript. A portion of this research was carried out at the Jet Propulsion Laboratory, California Institute of Technology, under a contract with the National Aeronautics and Space Administration (80NM0018D0004). This research used the NASA ADS and Simbad (CDS) databases. This work is based on observations made with the NASA/ESA/CSA James Webb Space Telescope. The data presented in this article were obtained from the Mikulski Archive for Space Telescopes (MAST) at the Space Telescope Science Institute, which is operated by the Association of Universities for Research in Astronomy, Inc., under NASA contract NAS 5-03127 for JWST. The specific observations analyzed can be accessed via \dataset[DOI: 10.17909/pv1h-ta47]{https://doi.org/10.17909/pv1h-ta47}. This research also uses data from IRSA and the included SOFIA science archive.

\end{acknowledgments}

\vspace{5mm}
\facilities{JWST(NIRCam), SOFIA(HAWC+), JCMT(SCUBA) }

\software{Astropy \citep{astropy:2013, astropy:2018, astropy:2022}} 

\software{DS9 \citep{ds92003}}

\software{IDL}

\appendix

\section{Notes on individual outflows}

\subsection{Outflow 1 (SH 2-68N)}
This outflow likely corresponds to the molecular (CO) outflow associated with SH 2-68N (S68N or J182948.1+011644; \citealt{Aso19, dunham15}), which in turn is part of the SMM 9 region \citep{Legoullec19,Tychoniec19}. The long wavelength emission has a PA of $\sim$135\degree. 

\subsection{Outflow 2 (S7)}
This is likely part of the blue lobe of the larger SMM 1 outflow, also known as S7 \citep{Herbst97, Caratti06}. Like several other sources in our sample, the IR morphology resembles the optical emission from an HH object with its umbrella-shaped bow shocks. This may be because the rotational H$_2$ emission dominates this source. While this is consistent with the lack of a clear driving source near the feature, we cannot rule out that this is a separate outflow coincident with the SMM 1 outflow based on the NIRCam image alone. Because of the ambiguity in interpretation, we classify this in the middle confidence bin.

\subsection{Outflow 3 (S68Nc)}
Outflow 3 (Figure \ref{fig:outflow1annot}) is identified as S9 in \cite{Herbst97}, and likely associated with the class 0 star S68Nc \citep{Aso19}. It is among the brightest outflows seen in the NIRCam field. The western extent of the outflow appears to be an isolated bow shock with the rest of the western lobe hidden by extinction. The eastern lobe is a prominent line of bow shocks, and then a significant break to a final bow shock on the eastern edge. Including all of these shocks increases the uncertainty of the PA, but the consistent arc suggests a slow precession. If we neglect the final Shock2E location, we find a mostly symmetric outflow with a steady precession rate of 5$\degree$ over 14$\arcsec$ of flow. Assuming a flow velocity of 100 km/s, that translates to a precession rate of $\sim$ 1$\degree$ per 57 yr. This is comparable to the rate of some other known precessing protostellar systems traced via outflow ejecta \citep[e.g.][]{Cunningham09}.

\subsection{Outflow 4 (OO Ser)}
This outflow is associated with the FUor candidate OO Ser \citep{Hodapp96}. It has a broad hourglass shape and relatively short extent in both directions, leading to a somewhat larger uncertainty in PA. 

\subsection{Outflow 5 (V370 Ser)}
This chain of knots points back to EC37/V370 Ser. While \cite{Hodapp12} was not able to measure a PA from H$_2$ emission near the source, the presence of this remote bow shock suggests a more edge-on orientation of the EC 37 system. \cite{Hodapp12} indicated that the knots to the west (MHO 2218) were likely associated with the nearer EC 37 system, and these knots (MHO 3245) are associated with OO Ser (based on the catalog from \citealt{Davis10}), the NIR bow shock directions suggest that OO Ser is ejecting the material in outflow 4.

\subsection{Outflow 6 (S68Nb2)}
This outflow is associated with the infrared-bright class 0/I source Serpens 7/S68Nb2 \citep{Gutermuth09, Aso19}. 

\subsection{Outflows 7 and 8}
Outflows 6 and 7 are likely associated with the SMM 1a and 1b binary, respectively, as their location and PAs match well with the CO outflows in \cite{Tychoniec19}. An alternate interpretation for outflow 7 has the driving source as the red protostar south of Serpens SMM1, known as EC 40 or SSTc2d J182949.6+011456 \citep{Gutermuth09}. In this work, we assume the latter scenario, because of the ALMA-derived kinematics of the dual outflow from the SMM1 binary \citep{Legoullec19,Tychoniec19}.

\subsection{Outflow 9}
This has in the past been associated with the SMM 1 outflow \citep{Caratti06}. However, we identify bipolar shapes that appear as bow shocks, which could be contrary to this interpretation. The orientation points back to SSTc2d J182950.5+011417 \citep{Harvey07}, although this would be newly identified as a driving source.

\subsection{Outflow 10 (EC 53)}

We interpret the driving source as the episodic flaring young protostar EC 53 \citep{Baek20}, driving a long chain of knots. The distance between the southernmost knots would suggest that mass loss episodes are $\sim$ 1000 yr apart, which is not consistent with the burst phase of EC 53 ($\sim$ 1.5 yr), although it is possible that the individual knots are each the result of a series of bursts. There is some evidence for precession as well.

\subsection{Outflow 11}

This outflow may be driven by Serpens 20. The center point is coincident with J182952.22+011547.4  \citep{Gutermuth09}, a young stellar object identified in the Spitzer catalog.

\subsection{Outflow 12}
It is unclear where in this morphologically complex flow the origin/driving source lies. There are a few options of nearby sources, including EC 55 \citep{Eiroa92}, which lies at the western terminus of the outflow as we characterize it in this catalog. For our purposes, we identify a knot of emission in the center that we ascribe to a previously unknown candidate driving source. While the well-characterized shape of the flow and clear directionality lends confidence in our identification of the outflow, a future proper motion observation would be required for confirmation.

\subsection{Outflow 13}
Outflow 13 is only detected via a single bow shock and no driving source is identified. The bow shock does not appear in F140M, which supports the shock interpretation, rather than scattered light off a pillar. The closest YSO is J18295354+0113051, a 2MASS source \citep{Cutri03} and detected with Gaia \citep{herczeg19}.

\subsection{Outflow 14 (Serpens 56)}
This outflow is associated with the nearby bright young star Serpens 56 \citep{Gutermuth09}. 

\subsection{Outflow 15}
Outflow 15 is only detected via a single bow shock and no driving source is identified. The bow shock does appear in F210M (and not F140M), which supports the shock interpretation. The closest YSO is J18295914+0114411, a 2MASS source \citep{Cutri03}.

\subsection{Outflow 16 (SMM3)}
This outflow is associated with the SMM3 protostar, which is not itself visible in any of the NIRCam bands, although it is well-detected by ALMA at 230 GHz, and by SCUBA at 450 $\mu$m \citep{Davis99}. The Class II YSO CK 8 is located (in projection) along the outflow, but does not appear to be interacting with it. The northern bow shock is very bright, and is partially saturated in F210M. There is a point source in the bow shock visible at F360M and F480M, but it is not clear if this is an unrelated embedded source.  

\subsection{Outflow 17}

Serpens 9, a radio (VLA) source and protostar to the east of the main cluster \citep{Bontemps96}, is well-aligned with the unipolar outflow, and we identify it as the driving source candidate.

\subsection{Outflow 18}

This outflow falls into our lowest confidence bin because of a non-visible driving source, and the somewhat disorganized shape of the knots to which we ascribe it, but does appear to be a symmetric bow shock around a submm source SMM11 \citep{Aso17a}.

\subsection{Outflow 19}
This outflow candidate falls into our lowest confidence bin. First, although we identify a potential driving source (Ser-emb 4E; \citealt{Enoch11}) based on the orientation of the bow shocks, there is no obvious nebulosity link between it and the outflow. Second, the outflow does not appear in the F210M band at all, suggesting it could have a different origin than shocked emission. Third, the tip resembles a cloud pillar, and sits near the highest extinction region in the southern region. 

\subsubsection{Outflow 20}

No apparent driving source is identified, but this object was previously noted as HH 459 \citep{Ziener99}. A candidate driving source is 2MASS J18300491+0114393.

\bibliography{outflows-nircam-2}{}
\bibliographystyle{aasjournal}

\end{document}